\documentclass[twocolumn,english,prl,floatfix,citeautoscript,nofootinbib,superscriptaddress]{revtex4}
\usepackage{amsbsy}
\usepackage{latexsym,epsfig,graphicx}
\usepackage{dcolumn}
\usepackage{subfigure}
\usepackage{comment}
\usepackage{color}
\usepackage[colorlinks,urlcolor=blue,citecolor=blue]{hyperref}
\usepackage{amstext}
\usepackage{amssymb}
\usepackage{setspace}

\begin{document}

\title{Supersymmetry-assisted high-fidelity ground state preparation of
single neutral atom in an optical tweezer}
\author{Xi-Wang Luo}
\affiliation{Department of Physics, The University of Texas at Dallas, Richardson, Texas
75080-3021, USA}
\author{Mark G. Raizen}
\affiliation{Center for Nonlinear Dynamics and Department of Physics, The University of
Texas at Austin, Austin, Texas 78712, USA}
\author{Chuanwei Zhang}
\email{chuanwei.zhang@utdallas.edu}
\affiliation{Department of Physics, The University of Texas at Dallas, Richardson, Texas
75080-3021, USA}

\begin{abstract}
Arrays of neutral-atom qubits in optical tweezers are a promising platform
for quantum computation. Despite experimental progress, a major roadblock
for realizing neutral atom quantum computation is the qubit initialization.
Here we propose that supersymmetry---a theoretical framework developed in
particle physics---can be used for ultra-high fidelity initialization of
neutral-atom qubits. We show that a single atom can be deterministically
prepared in the vibrational ground state of an optical tweezer by
adiabatically extracting all excited atoms to a supersymmetric auxiliary
tweezer. The scheme works for both bosonic and fermionic atom qubits trapped
in realistic Gaussian optical tweezers and may pave the way for realizing
large scale quantum computation, simulation and information processing with
neutral atoms.
\end{abstract}

\maketitle


\section{Introduction}

Neutral atoms trapped in optical tweezer arrays 
have emerged as a promising candidate for quantum computation and simulation
\cite%
{Schlosser2001sub,RevModPhys.82.2313,RevModPhys.86.153,PhysToday.70.44,Science.357.995}
due to their attractive features such as identical qubits, large scalability
through atom-by-atom assemblers \cite%
{PhysRevA.70.040302,Science.354.1024,ncomms13317,science.aah3778,OE.24.009816,Nature561.79,PhysRevLett.122.203601}%
, and high precision control and measurement. For neutral-atom qubits,
high-fidelity single qubit gates have been realized using microwave or
two-photon Raman transitions \cite%
{PhysRevLett.96.063001,PhysRevA.74.042316,nature09827,PhysRevLett.93.150501,science.aaf2581,PhysRevA.77.052309,PhysRevLett.121.240501}%
. Two-qubit gates have been realized using short-range collisions or
long-range Rydberg interactions \cite%
{Nature527.208,PhysRevLett.85.2208,PhysRevLett.104.010503,PhysRevLett.104.010502,nphys3487,nature24622,PhysRevLett.121.123603,PhysRevX.8.021070,Saffman2019}%
, with significantly improved gate fidelity in recent years.

Significant experimental progress has been made on high-fidelity neutral
atom qubit initialization that requires deterministic preparation of single
atom on the vibrational ground state of an optical tweezer, but major
obstacles still exist. For bosonic atoms, interaction blockade and
single-atom rapid imaging allow the deterministic preparation of a single
atom in an optical tweezer, and defect-free atom arrays with up to tens of
single atoms have been demonstrated by rearranging the occupied tweezers~%
\cite{Science.354.1024,ncomms13317,science.aah3778,OE.24.009816,Nature561.79}%
. However, atoms in the tweezers are subject to imaging heating and the
experimental ground-state cooling is far from perfect due to photon recoil
in sideband cooling\ \cite%
{PhysRevX.2.041014,PhysRevLett.110.133001,PhysRevA.97.063423,PhysRevX.9.021039,PhysRevX.8.041055,PhysRevX.8.041054,PhysRevLett.122.143002}%
. For fermionic atoms, high-fidelity preparation of a few atoms is possible
through the method of trap deformation~\cite%
{PhysRevA.80.030302,science.1201351,PhysRevLett.115.215301,science.1240516}.
However, to obtain a single fermion ground state, the trap need be tilted
and ramped down to an extremely low depth to spill the excess atoms, making
the process very sensitive to external potential noises and requiring a long
trap-deforming time to avoid heating. For both bosons and fermions, the
fidelity to prepare a single atom in the ground state of a tweezer is $\sim
90\%$\ \cite%
{PhysRevX.2.041014,PhysRevA.97.063423,PhysRevLett.110.133001,PhysRevX.9.021039,science.1201351}
in realistic experiments.

Supersymmetry was first introduced within the context of particle physics
and became one possible solution to many important problems in high-energy
physics~\cite{Supersymmetry2015}. Though supersymmetry remains to be
observed in particle physics, it has found applications in areas including
condensed matter physics, cold atoms and optics~\cite%
{Supersymmetry1995Cooper,PhysRevLett.100.090404,PhysRevLett.110.233902,ncomms4698,OL.43.004927,LiangFeng2019Super,science.aav5103}%
.

In this Letter, we propose a supersymmetry-based scheme to achieve
ultra-high fidelity single atom ground state preparation in an optical
tweezer through adiabatically extracting excited atoms to its supersymmetric
partner, an auxiliary tweezer. Specifically, the eigenstates of two tweezers
(except the main tweezer ground state) are pairwise related to one another,
yielding supersymmetry (either exact or approximate). For bosons, we can
prepare a sideband-cooled single atom \cite%
{PhysRevX.2.041014,PhysRevLett.110.133001,PhysRevA.97.063423,PhysRevX.9.021039,PhysRevX.8.041055,PhysRevX.8.041054,PhysRevLett.122.143002}
and transfer its excited components to the supersymmetric auxiliary trap,
followed by postselecting the measurement result with an empty auxiliary
tweezer (\textit{i.e.}, the single atom stays on the ground state of the
original tweezer). For fermions, we start from a few atoms \cite%
{science.1201351} occupying the low-lying states and transfer all excited
atoms to the supersymmetric auxiliary tweezer. We consider realistic
Gaussian optical tweezers and show that ultra-high fidelity ground state
preparation can be achieved in a short time interval for both bosons and
fermions. In such qubit initialization process, supersymmetry plays an
essential role for simultaneously extracting all excited components or atoms
from the main tweezer to the auxiliary tweezer.

\section{Supersymmetry}

\begin{figure}[t]
\includegraphics[width=0.95\linewidth]{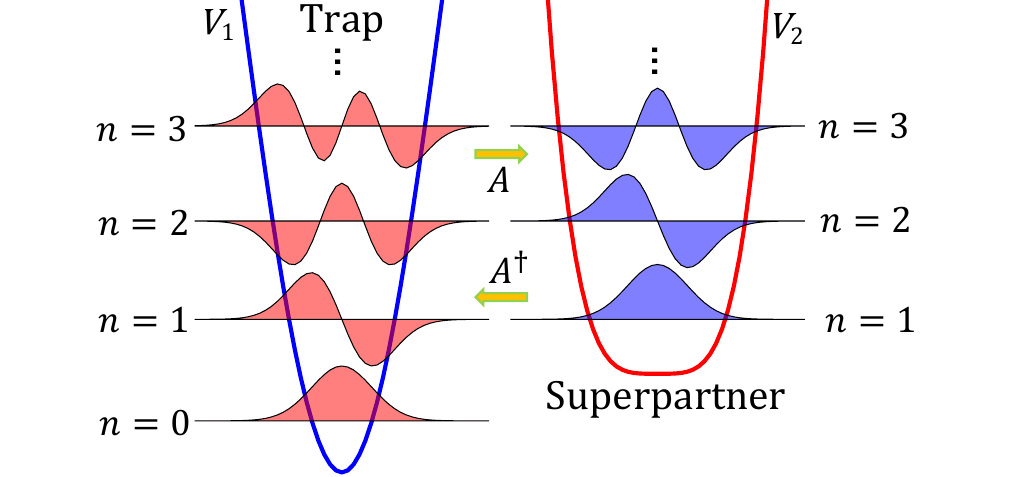}
\caption{Schematic illustration of a trap potential and its supersymmetric
partner. Except the ground state, all eigenvalues of the trap are exactly
matched to those of its superpartner. The corresponding eigenstates are
related through the action of $A$ and $A^{\dag }$.}
\label{fig:susy}
\end{figure}

In quantum mechanics, supersymmetry theory involves a pair of partner
Hamiltonians such that for every eigenstate $|\varphi _{1,n}\rangle$ (except
the $n=0$ ground state) of one Hamiltonian $H_{1}$, its partner Hamiltonian $%
H_{2}$ has a corresponding eigenstate $|\varphi _{2,n}\rangle$ with the same
energy~\cite{Supersymmetry1995Cooper}. This can be established by
factorizing the Hamiltonian in terms of two operators $A$ and $A^{\dagger }$
\begin{equation}
H_{1}=A^{\dagger }A,H_{2}=AA^{\dagger },  \label{Eq:H12}
\end{equation}%
%
%
%
%
%
which are isospectral 
with eigenstates (non-normalized) pairwise related to one another through $%
|\varphi _{2,n}\rangle =A|\varphi _{1,n}\rangle $ and $|\varphi
_{1,n}\rangle =A^{\dagger }|\varphi _{2,n}\rangle $. If the ground state of $%
H_{1}$ is annihilated by $A$, i.e., $A|\varphi _{1,0}\rangle =0$, then it
does not have a corresponding state in $H_{2}$,
leading to exact supersymmetry between the two Hamiltonians, as
schematically illustrated in Fig.~\ref{fig:susy}. 

For the nonrelativistic Schr\"{o}dinger problems, one can always identify
two supersymmetric potentials, $V_{1}(x)$ and the superpartner $V_{2}(x)$,
that are entirely isospectral except for the ground state of $V_{1}(x)$.
Here we are interested in neutral atoms trapped in optical tweezers, and
only low energy bound states are relevant. We consider a deep optical
tweezer with $N_{\text{b}}$ bound states that are filled with $N_{\text{a}}$
non-interacting neutral atoms ($N_{\text{a}}\ll N_{\text{b}}$). Only the
first $N$ bound states (with $N\ll N_{\text{b}}$) are relevant if the system
is pre-cooled to a low temperature.
Here we still call $V_{2}(x)$ the superpartner tweezer of $V_{1}(x)$ if the
first $N$ bound-state energies of $V_{1}(x)$\ (except the ground state) are
exactly matched by the first $N-1$ bound-state energies of $V_{2}(x)$.

\section{Adiabatic extraction}

Although our scheme can be applied to any dimension, we will first limit our
analysis to one spatial dimension to simplify the calculation. We assume
significantly strong trapping along transverse directions, where only the
ground transverse state is occupied. We consider a main optical tweezer $%
V_{1}(x)$ with non-interacting atoms populating only the first $N$ bound
states (\textit{i.e.}, the populations on higher energy states are
negligible), and introduce an auxiliary empty tweezer $V_{2}(x)$ that is the
superpartner of $V_{1}(x)$. Within the subspace spanned by the first $N$
bound states, the Hamiltonians read $H_{1}=\sum_{n=0}^{N}E_{1,n}|\varphi
_{1,n}\rangle \langle \varphi _{1,n}|$, $H_{2}=\sum_{n=1}^{N}E_{2,n}|\varphi
_{2,n}\rangle \langle \varphi _{2,n}|$ with $E_{1,n}=E_{2,n}$ for $n\geq 1$.

\begin{figure}[t]
\includegraphics[width=0.95 \linewidth]{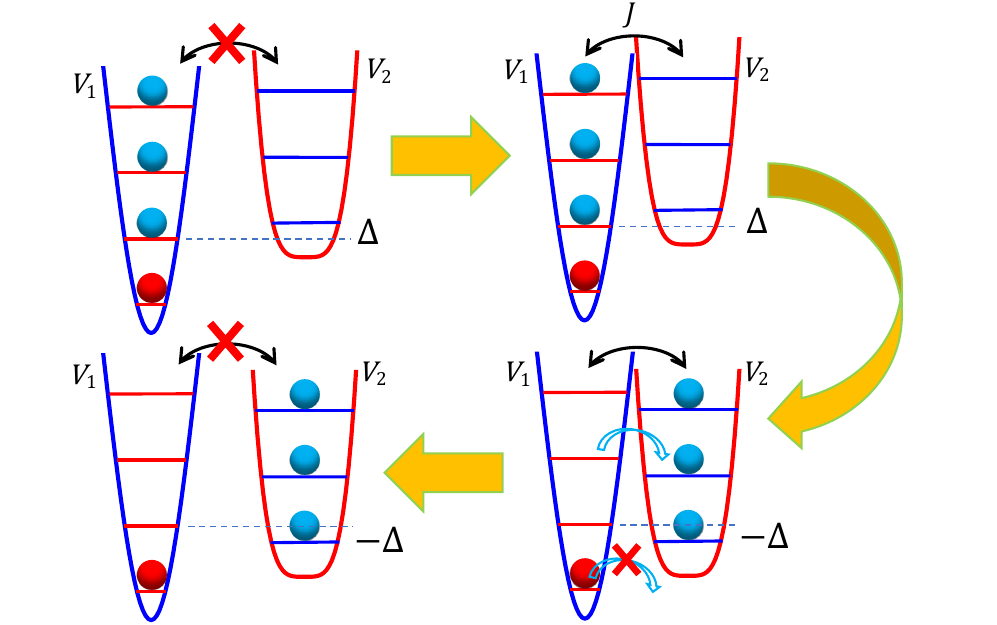}
\caption{Schematic illustration of adiabatic extraction of all excited atoms
(or components) in the optical tweezer using the superpartner as an
auxiliary tweezer. $J$ and $\Delta $ are the tunneling amplitude and
detuning between two tweezers.}
\label{fig:adiabatic}
\end{figure}

The adiabatic atom extraction is performed as follows: (i) The auxiliary
tweezer $V_{2}(x)$ is deformed such that its eigenenergies are increased by $%
\Delta $. (ii) $V_{2}(x)$ is transported toward the main tweezer $V_{1}(x)$
adiabatically, and the bound state $|\varphi _{1,n}\rangle $ is coupled with
its counterpart $|\varphi _{2,n}\rangle $ for $n\geq 1$. (iii) $V_{2}(x)$ is
adiabatically deformed to decrease its eigenenergies by $-2\Delta $ and then
transported away from the main tweezer $V_{1}(x)$. (iv) The original empty $%
V_{2}(x)$ is restored for next extraction. After such an adiabatic process,
all atoms (or atom components) in the excited states of the main tweezer are
transported to the auxiliary tweezer, while atom (component) in the ground
state $|\varphi _{1,0}\rangle $ is unaffected, as illustrated in Fig.~\ref%
{fig:adiabatic}. As a result, the remaining atom (component) in the main
tweezer is prepared in the vibrational ground state. Here the pairwise
energy levels due to supersymmetry play a central role for the atom
extraction. 
Steps (i) and (iv) can be done very fast, we will focus on steps (ii) and
(iii) in the following.

The time-dependent effective Hamiltonian can be written as $H_{\text{tot}%
}(t)=H_{1}+H_{2}+H_{\text{int}}(t)$ (see Appendix A) with%
\begin{equation}
H_{\text{int}}(t)=\sum_{n=1}^{N}\frac{\Delta _{n}}{2}|\varphi _{2,n}\rangle
\langle \varphi _{2,n}|+J_{n}(t)|\varphi _{1,n}\rangle \langle \varphi
_{2,n}|+h.c.,
\end{equation}
where we have neglected the far-off-resonance couplings which do not affect
the adiabatic extraction. In fact, the adiabatic process is robust against
perturbations, and the deformation and transport of the auxiliary tweezer
are very flexible. The only requirement is that $|\Delta _{n}|$, $|J_{n}|$
are small compared to the level splitting $|E_{1,n}-E_{1,n\pm 1}|$ during
the adiabatic process, such that all energy levels are gapped (see Appendix
B). Furthermore, even when $V_{2}(x)$ is not initialized as the exact
superpartner of $V_{1}(x)$ with $E_{1,n}\neq E_{2,n}$, we can still extract
all excited atoms if we have approximate supersymmetry (i.e., the symmetry
breaking is weak with 
$|E_{1,n}-E_{2,n}|\ll 
|E_{1,n}-E_{1,n\pm 1}|$) (see Appendix C). The extraction would fail if the
supersymmetry is strongly broken.

\begin{figure}[t]
\includegraphics[width=1.0\linewidth]{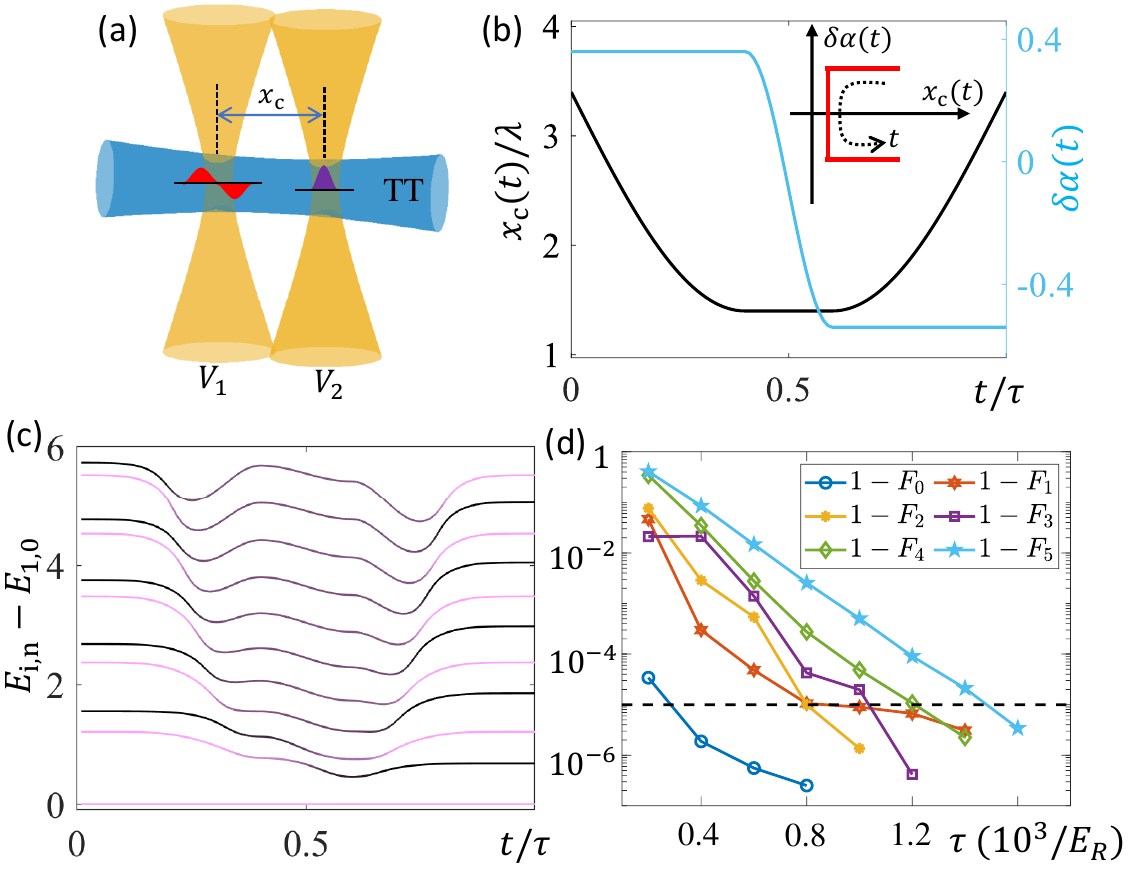}
\caption{(a) Gaussian optical tweezer and its superpartner with additional
transverse trap (TT). (b) Trap depth and center of the auxiliary
(superpartner) tweezer as functions of time during the adiabatic extraction.
The inset shows the adiabatic path. (c) Total eigenspectrum of the system
along the path in (b). Colors correspond to the population probability in
the two tweezers, with pink (light gray) and black representing full population in the main
and auxiliary tweezer, respectively. (d) The fidelity $F_{n}$ as a function of adiabatic
duration length $\protect\tau $.}
\label{fig:fidelity}
\end{figure}

\section{Physical realization}

Although our proposal does not rely on the specific shape of the tweezer,
here we consider a realistic Gaussian trap function $V_{1}(x)=\alpha
_{1}e^{-2x^{2}/w_{0}^{2}}$ (see Fig.~\ref{fig:fidelity}a), where $w_{0}$ is
the width and $\alpha _{1}$ is the trap depth. As an example, we choose a
typical width $w_{0}=1$ $\mu $m~\cite{Science.354.1024} and the trapping
wavelength $\lambda =810$nm~\cite{Science.354.1024}, and use the recoil
momentum $k_{\text{R}}=\frac{2\pi }{\lambda }$ and energy $E_{\text{R}}=%
\frac{\hbar ^{2}k_{\text{R}}^{2}}{2m}$ as the units (with $m$ the atom
mass). For a deep Gaussian trap, the low energy dynamics is approximately
characterized by a harmonic oscillator with equal energy splitting. The
superpartner of a harmonic trap can be easily obtained by a constant shift
of the trapping potential that equals to the trapping frequency. Therefore,
slight change in Gaussian trap depth leads to an approximate superpartner
trap $V_{2}(x,t)=[\alpha _{2}+\delta \alpha (t)]e^{-2[x-x_{\text{c}%
}(t)]^{2}/w_{0}^{2}}$. With proper choice of $\alpha _{1}$ and $\alpha _{2}$
(e.g., $\alpha _{1}=-12E_{\text{R}}$ and $\alpha _{2}=-10.76E_{\text{R}}$),
the energy levels (we consider $N=5$ here) of two optical tweezers are
paired except for the ground state of $V_{1}$ (see Appendix C). The
extraction is realized by adiabatically tuning the depth $\delta \alpha (t)$
and center $x_{\text{c}}(t)$ of $V_{2}$, as shown in Fig.~\ref{fig:fidelity}%
b. The creation and manipulation of controlled optical tweezers can be
accomplished with an electro-optic deflector which toggles between two
voltages on a sufficiently-fast time scale so that the atoms experience a
time-averaged effective potential~\cite%
{PhysRevLett.86.1514,PhysRevLett.86.1518}. Merging and separating the
supersymmetric tweezer pairs could be done by a programmed sequence of
voltages that are applied to an electro-optic deflector.

In Fig.~\ref{fig:fidelity}c, we plot the spectrum of the system during the
adiabatic process along the path in Fig.~\ref{fig:fidelity}b, which is
obtained by solving the real-space Schr\"{o}dinger equation $\left[ -\frac{%
\hbar ^{2}\partial _{x}^{2}}{2m}+V_{1}(x,t)+V_{2}(x,t)\right] |\varphi
\rangle =E|\varphi \rangle $. We see the spectrum is gapped all the time,
while the eigenstates in the two tweezers exchange except for the ground
state of $V_{1}$. In principle, the extraction fidelity can achieve 100\%
for sufficient long adiabatic interval $\tau $. Assuming an atom stays in
state $|\varphi _{1,n}\rangle $ at time $t=0$, we define the fidelity $F_{n}$
as the probability to find the atom at time $t=\tau $ on the ground state $%
|\varphi _{1,0}\rangle $ of the main tweezer for $n=0$ or in the auxiliary
tweezer for $n>0$. In Fig.~\ref{fig:fidelity}d, we show the fidelities as
functions of $\tau $ obtained from numerical simulating the time-dependent
real-space Schr\"{o}dinger equation. The extraction process is a multistate
Landau--Zener problem (see Appendix B), and there are couplings between
different eigenstates when $\tau $ is small, yielding fidelities that are
far below $1$. The fidelity $F_{n}$ can be close to $1$ at a larger $\tau $.
For a realistic shallow trap $\alpha _{1}=-12E_{R}$, the fidelity can be up
to $F_{n}\gtrsim 1-10^{-5}$ with a short adiabatic interval $\tau \gtrsim 7$%
ms ($\tau \gtrsim 70$ms) for Li (Rb) atoms, which can be improved further by
optimizing the adiabatic loop or using deeper tweezers.

\begin{figure}[t]
\includegraphics[width=0.9\linewidth]{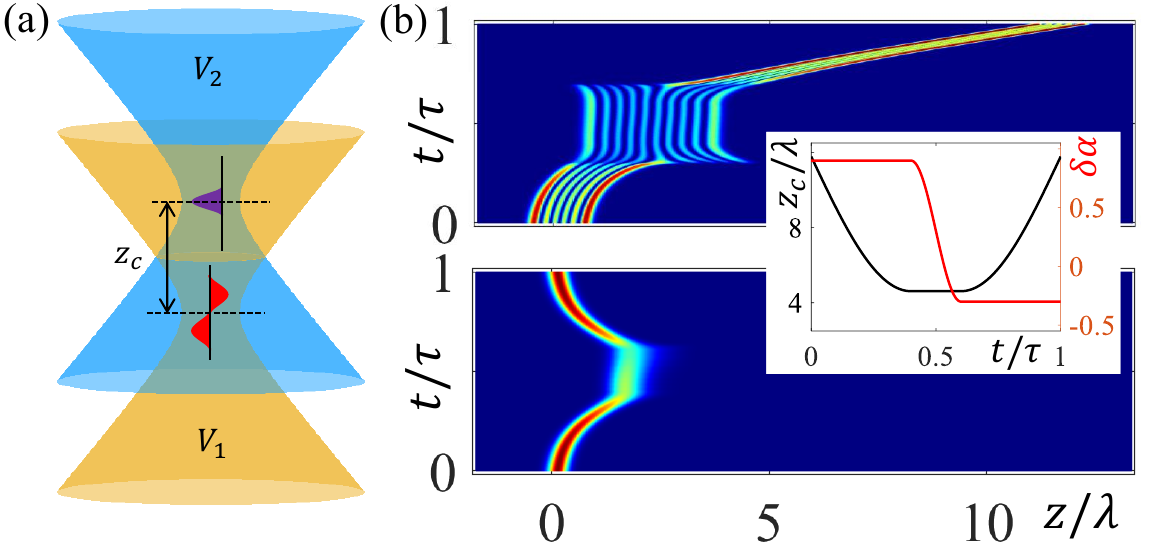}
\caption{(a) Two tweezers with longitudinal relative shift $z_{c}$. (b) Atom
distributions during the extraction for the ground state (bottom) and 5-th
excited state (top) obtained from a full 3D simulation of the Schr\"{o}%
dinger equation. The inset shows $z_{c}(t)$ and $\protect\delta \protect%
\alpha (t)$ during the adiabatic process with $\protect\alpha _{1}=200E_{R}$%
, $\protect\alpha _{2}=198.8E_{R}$, $w_{0}=0.9\protect\mu $m. Other
parameters are the same as in Fig.~\protect\ref{fig:fidelity}. The results
for other excited state are similar. The fidelities are $F_{n}\sim 1-10^{-4}$
(up to $n=5$) with adiabatic interval $\protect\tau \sim 20$ms ($\protect%
\tau \sim 200$ms) for Li (Rb) atoms. }
\label{fig:3Dtw}
\end{figure}

We now consider the tweezer realized by a single strong and tightly focused
Gaussian beam without additional transverse trapping (i.e., 3D). The trap is
given by $V_{1}(\mathbf{r})=\alpha _{1}\frac{w_{0}^{2}}{w_{z}^{2}}\exp
[-2(x^{2}+y^{2})/w_{z}^{2}]$ with the spot size $w_{z}=w_{0}\sqrt{1+\frac{%
z^{2}}{z_{R}^{2}}}$ and Rayleigh range $z_{R}=\frac{\pi w_{0}^{2}}{\lambda }$%
. In this case, the tight transverse trapping, realized by the Gaussian
tweezer itself, is much stronger than the longitudinal trapping. Therefore,
atoms would stay in the transverse ground state with high-probability after
sideband cooling (laser culling) for bosons (fermions)~\cite%
{PhysRevX.2.041014,PhysRevLett.110.133001,science.1201351}, and the overall
fidelity is mainly limited by the residual excitations in the longitudinal
direction. We may 1) slowly bring the auxiliary tweezer to the main tweezer
with their beam waists largely separated by $z_{c}$ along the longitudinal
direction (see Fig.~\ref{fig:3Dtw}a); 2) tune $z_{c}$ and $\delta \alpha $
along the adiabatic path (see the inset of Fig.~\ref{fig:3Dtw}b); 3) slowly
move the auxiliary tweezer away from the main tweezer. Steps 1) and 3) can
be done with high fidelity due to large separation of the two tweezers. Here
we focus on step 2) and solve the 3D Schr\"{o}dinger equation numerically.
We find that the extract fidelity can be up to $F_{n}\sim 1-10^{-4}$ with
proper choice of parameters, as shown in Fig.~\ref{fig:3Dtw}b. A longer
extraction time is needed to achieve higher fidelity. Here the transverse
and longitudinal modes do not mix~\cite{J.Mod.Opt.54.1619} for the two
tweezers shown in Fig.~\ref{fig:3Dtw}a.

\section{Boson qubit initialization}

We assume an initial low-temperature single-atom state (i.e., $N_{a}=1$).
This is because, more than one atom may be left on the ground state after
the extraction for $N_{a}>1$ non-interacting bosons, and the energy level
coupling with the auxiliary tweezer would be strongly modified for
interacting bosons. Fortunately, deterministic preparation of a single boson
in an optical tweezer has been realized through single atom imaging and the
atom can be further sideband cooled with ground state population $\sim 90\%$~%
\cite{Science.354.1024,PhysRevX.2.041014,PhysRevA.97.063423}. Assuming a
thermal population distribution, the total probability to find the atom on $%
n>N$ states is $<10^{-5}$ for $N=5$. After the supersymmetric adiabatic
extraction process, all excited components of the atom are transferred to
the auxiliary tweezer, in which the atom number is measured. If one atom is
detected in the auxiliary tweezer, we discard the atom qubit in the main
tweezer and the process fails. If no atom is probed in the auxiliary
tweezer, the atom must be on the ground state of the main tweezer, therefore
we keep the atom qubit in the main tweezer. The process is successful and we
know with 100\% probability that there is a single atom on the ground state
of the main tweezer. Such postselection measurement leads to deterministic
preparation of a single atom in the main tweezer with a total ground-state
fidelity $\geq 1-\sum_{n>0}\frac{P_{n}(1-F_{n})}{P_{0}F_{0}}\geq 1-10^{-5}$ (%
$P_{n}$ is the $n$-th state occupation probability). Here postselection is
to condition a probability space upon the occurrence of a given event, and
the fidelity is defined as the probability to find the atom in the ground
state of the main tweezer upon the occurrence of an empty auxiliary tweezer.
Notice that the success probability is $P_{0}F_{0}$ while the probability to
find an empty auxiliary tweezer is $P_{0}F_{0}+\sum_{n>0}P_{n}(1-F_{n})$.
Therefore, the fidelity is $P_{0}F_{0}[P_{0}F_{0}+%
\sum_{n>0}P_{n}(1-F_{n})]^{-1}\geq 1-\sum_{n>0}\frac{P_{n}(1-F_{n})}{%
P_{0}F_{0}}$. The detection of the auxiliary tweezer can be done by
single-atom-resolved fluorescence imaging technique~\cite%
{nature09378,nature09827,RevModPhys.82.2313}, where a practical issue is
that the resulting scattered resonant light may be absorbed by other qubits,
degrading their fidelity (note that fermions do not need resonant detection
during preparation, which is an advantage, see below). The resonant
scattering light could be avoided by first transferring the
auxiliary-tweezer atom to another hyperfine state (e.g., $F=2$ state of $%
^{87}$Rb) with $\sim $GHz energy splitting \cite{PhysRevA.74.042316}, where
the imaging laser (focused on the auxiliary tweezer) is far-off-resonance
with the main-tweezer atoms in other qubits, thus would not disturb their
states.


\section{Fermion qubit initialization}

For fermions, 
it is more convenient to start from several atoms (e.g., $N_{a}=4,5$)
distributed on the low-lying energy levels in the tweezer (see Appendix D),
then we apply the supersymmetric adiabatic atom extraction process to obtain
a single ground-state fermion. Note that no postselection measurement of the
auxiliary tweezer is needed for fermionic qubits. We consider spin-polarized
fermions with negligible interactions due to the antisymmetric wavefunction.
The preparation fidelity is $\geq P_{0}\prod_{n=0}^{N_{a}-1}F_{n}$.
If one loads fermions from a reservoir with typical temperature $T/T_{F}=0.5$
to a tweezer with depth $5k_{B}T_{F}$, one obtains $P_{0}>1-10^{-5}$~\cite%
{PhysRevA.63.033603}. The total fidelity can be up to $\sim 1-10^{-5}$. To
obtain a tweezer with $N_{a}$ low-energy atoms, one can first load a large
number of fermions from a reservoir and spill excess highly excited atoms by
varying the depth of the tweezer and the strength of a magnetic field
gradient \cite{PhysRevA.80.030302,science.1201351}. For a large $N_{a}$
(e.g., $N_{a}=4,5$), the tweezer depth remains much higher than the ground
state energy whose occupation is hardly affected during the spilling
process. 
Moreover, imperfect spilling that change $N_{a}$ by $\pm 1$ or $\pm 2$ do
not affect our high-fidelity preparation as long as the ground state is
occupied with a high probability.

It is also possible to prepare a single fermion to the ground state based on
the spilling method in~\cite{PhysRevA.80.030302,science.1201351}, however,
the overall fidelity is very limited in realistic experiments. This is
because the trap need be tilted and ramped down to an extremely low depth,
which not only makes the spilling process very sensitive to potential noises
(induced by fluctuations in laser intensity and magnetic field), but also
requires a long trap-deforming time to avoid heating.

\section{Discussion}

The fidelity might be slightly suppressed by the common heating sources
existing in the system. The atom heating due to off-resonance light
scattering (the rate is $\sim \frac{V}{\Delta _{\text{e}}}\Gamma $ with $V$
the trap depth and $\Gamma $ the damping rate of the excited state) is
negligible for a large detuning $\Delta _{\text{e}}$ to the excited state~%
\cite{AdvAtMolOptPhys.42.95}. Intensity fluctuations of the trapping lasers
can be very weak and dominated by low-frequency (much smaller than the
trapping frequency) noise using intensity stability techniques~\cite%
{OptLett.34.2912, OptLett.42.755, PhysRevLett.121.173601}, therefore the
fluctuations have very minor effects in the adiabatic process~\cite%
{PhysRevA.56.R1095,PhysRevA.80.032307}. The fluctuation effects can be
further suppressed by using the same laser source for both the main and
auxiliary tweezers. Background gas collision-limited lifetime is about 10
seconds~\cite{Science.354.1024,ncomms13317,science.aah3778}, which is much
longer than the adiabatic duration. The adiabatic transfer time for Li and
Rb atoms is about 7 and 70 milliseconds, and both can be further improved
using deeper dipole traps (the shallow trap considered above has a trapping
frequency only about several kHz). Here the adiabatic duration can be at the
same order as the sideband cooling (i.e., several milliseconds)~\cite%
{PhysRevX.2.041014}. Therefore, our scheme can lead to high-fidelity and
fast qubit preparation even in the presence of these common heating sources
in realistic experiments. We want to point out that, for boson qubit
preparation based on sideband cooling only, one could improve the fidelity
by decreasing the Raman beam power and increasing the detuning with the
excited state~\cite{PhysRevLett.75.4011}, but the cooling time will also
increase significantly, which may enhance the heating from other sources.
The cooling fidelity may also be improved using very deep traps, which is,
however, quite limited for neutral atoms (typical tweezer trapping frequency
ranges from several kHz to several 10kHz, comparing to 10MHz for ion traps).
More importantly, our adiabatic extraction scheme applies to both fermion
and boson qubit preparation, while the sideband cooling cannot be used for
fermion qubit preparation with many atoms in the tweezer initially.

Combined with the capability of rearranging tweezers, our method can
initialize a large array of neutral atom (bosonic or fermionic) qubits to
the vibrational ground state. In addition to quantum computation, such
ground state single-atom tweezers can be used as building blocks for
generating entangled states such as Dicke and NOON states~\cite%
{RevModPhys.90.035005} that are useful for high precision quantum metrology
beyond the standard quantum limit $1/\sqrt{N}$.

\begin{figure}[tb]
\includegraphics[width=1.0\linewidth]{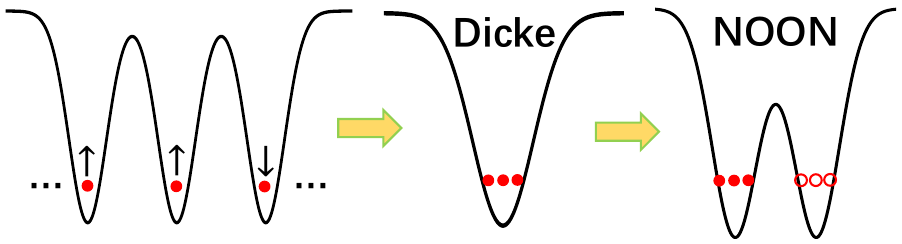}
\caption{Scheme to generate nonclassical entangled Dicke and NOON states by
merging and splitting many single-atom tweezers. The spin states could be
atomic hyperfine states.}
\label{fig:NOON}
\end{figure}

The Dicke state is a symmetrized spin state with total spin $J$ and
z-component $m_{z}$, corresponding to a two-mode Fock state with $J\pm m_{z}$
atoms in spin up and down. Such Dicke state can be realized by merging $2J$
single-atom optical tweezers, with $J\pm m_{z}$ tweezers containing spin-up
and down atoms (see Fig.~\ref{fig:NOON}). The repulsive interaction is
turned on to ensure $2J$ atoms remaining on the ground state \cite%
{PhysRevLett.91.010402} during the adiabatic evolution. Here the many-body
energy gap during the adiabatic merging is roughly given by the smaller one
of two energy scales: the interaction energy $E_{\text{int}}$ (interaction
between two atoms in one tweezer) and excited state energy $E_{\text{e}}$
(when the barriers between neighboring tweezers vanish). For typical
tweezers and atom scattering lengths, $E_{\text{int}}$ can be up to several
tens Hz, and $E_{\text{e}}\sim \frac{E_{R}}{4J^{2}}$ is around a hundred Hz
for $J=10$ (i.e., 20 atoms), leading to the adiabatic merging time $\sim $
10ms.

With $2J$ atoms on the ground state of one tweezer, we can slowly switch the
repulsive interaction to attractive, then split the tweezer into two
identical tweezers (see Fig.~\ref{fig:NOON}), generating a NOON state (i.e.,
a coherent superposition of all particles in the left or right tweezer)~\cite%
{PhysRevLett.91.010402}. If the interaction energy is smaller than the
single tweezer trapping frequency, even a sudden switching of the
interaction would not excite the system~\cite{PhysRevLett.91.010402}, which
is still satisfied with 20 atoms in one tweezer. During this splitting, the
many-body gap is enhanced by $J$ times comparing to the merging process,
thus can be done much faster. Both Dicke and NOON states can yield
measurement precision scaling as the Heisenberg limit $\sim 1/N$~\cite%
{RevModPhys.90.035005}.

Finally, the ability of generating a few-atom Fock state in the tweezer
provides a new platform for studying few-body physics with the fixed atom
number. For instance, by tuning the interaction through Feshbach resonance,
it is possible to study the universality of Efimov trimer and other
multi-body bound states~\cite%
{Science.326.1683,Rep.Prog.Phys.75.046401,Rep.Prog.Phys.80.056001,RevModPhys.89.035006}%
.

\section{Conclusion}

In summary, we propose a method to deterministically prepare a single atom
to the vibrational ground state of an optical tweezer with high fidelity,
using a supersymmetric auxiliary tweezer. The supersymmetry is crucial for
tweezer geometry design and plays a central role for extracting excited
atoms. The scheme is built upon recent experimental progress on single atom
preparation and sideband cooling, and applies to both fermionic and bosonic
atom qubits. It addresses one major roadblock for realizing high-fidelity
neutral atom qubit initialization, therefore may pave the way for the
experimental realization of intermediate-scale neutral atom quantum
computation and simulation. Our proposed qubit initialization can also be
used to generate nonclassical quantum states which may find applications in
other fields such as high precision measurement and quantum sensors.

\begin{acknowledgments}
\textbf{Acknowledgements}: XWL and CZ are supported by AFOSR
(FA9550-16-1-0387, FA9550-20-1-0220), NSF (PHY-1806227), and ARO
(W911NF-17-1-0128). Part of C.Z. work was performed at the Aspen Center for
Physics, which is supported by National Science Foundation grant PHY-1607611.
\end{acknowledgments}

\section{Appendix A: Effective Hamiltonian}

Here we show how the simple model of Eq.~(2) in the main text can be used to
describe the adiabatic process. We first consider the 1D case, and a double
well trap $V(x,t)=V_{1}(x)+V_{2}(x,t)$ is formed by the two tweezers at time
$t$ during the adiabatic process. The low-energy local modes $|\varphi
_{1,n}(t)\rangle $ of the left well (corresponding to the main tweezer) can
be approximately obtained by solving a harmonic oscillator with curvature
and center determined by the left trap minimum, similarly for the right well
(auxiliary tweezer) $|\varphi _{2,n}(t)\rangle $. Then the coupling reads
\begin{equation}
J_{n,m}(t)=\int \varphi _{1,n}^{\ast }(x,t)[-\frac{\hbar ^{2}\partial
_{x}^{2}}{2m}+V(x,t)]\varphi _{2,m}(x,t)dx.
\end{equation}%
The integral $J_{n,m}$ is nonzero even when the wave functions $\varphi
_{1,n}(x,t)$ and $\varphi _{2,m}(x,t)$ have opposite parity, since their
parity-symmetry centers are different. As the two tweezers approach each
other, the energy levels may also shift slightly, since the curvature of the
main (auxiliary) tweezer would be modified by the auxiliary (main) tweezer.
The full Hamiltonian reads $H_{\text{tot}}(t)=H_{1}(t)+H_{2}(t)+H_{\text{int}%
}(t)$ with
\begin{equation}
H_{i}(t)=\sum_{n=1}^{N}E_{1,n}(t)|\varphi _{i,n}\rangle \langle \varphi
_{i,n}|,  \label{Eq:H12S}
\end{equation}
and
\begin{equation}
H_{\text{int}}(t)=\sum_{n=1}^{N}\frac{\Delta _{n}}{2}|\varphi _{2,n}\rangle
\langle \varphi _{2,n}|+J_{n}|\varphi _{1,n}\rangle \langle \varphi
_{2,n}|+h.c.  \label{Eq:HintS}
\end{equation}%
The off-resonance coupling terms $J_{n,m}|\varphi _{1,n}\rangle \langle
\varphi _{2,m}|+h.c.$ with $m\neq n$ are neglected.

The effective Hamiltonian is similar for the 3D case if we focus on the
transverse ground-state subspace, which reads
\begin{equation}
J_{n,m}(t)=\int \varphi _{1,n}^{\ast }(\mathbf{r},t)[-\frac{\hbar ^{2}%
\mathbf{\nabla }^{2}}{2m}+V(\mathbf{r},t)]\varphi _{2,m}(\mathbf{r},t)dx.
\end{equation}%
Here $\varphi _{1,n}(\mathbf{r},t)$ is the wave function of the $n$-th
longitudinal mode in the transverse ground state. Notice that, $\varphi
_{1,n}(\mathbf{r},t)$ and $\varphi _{2,m}(\mathbf{r},t)$ have the same
transverse parity-symmetry center. They stay in the ground transverse state
and have the same transverse parity, therefore can couple with each other.
The two tweezers are parallel with each other, and they have the same
optical axis as shown in Fig.~4a in the main text; therefore, transverse and
longitudinal modes do not mix for our system. In fact, the extraction of
excited longitudinal modes is independent from the transverse state of the
atoms in the main tweezer. Generally, the transverse modes are more
confined, which typically have six (or more) times larger trapping frequency
than the longitudinal modes. Therefore, high-fidelity transverse ground
state can be obtained by sideband cooling for bosons or by spilling for
fermions, with occupation only on the first $N$ longitudinal modes.

\section{Appendix B: Adiabatic condition}

The simple model given by Eqs.~\ref{Eq:H12S} and \ref{Eq:HintS} describes
many independent two-level Landau-Zener processes. We have assumed that the
detuning $\Delta _{n}$ and coupling $J_{n}$ are small comparing to the
tweezer energy level splitting $\omega _{n}=E_{1,n+1}-E_{1,n}$, so the
off-resonance couplings are neglected. The instantaneous eigenenergy levels
and eigenstates of the $n$-th Landau-Zener pair are $\varepsilon _{n,\pm
}=E_{1,n}\pm \sqrt{J_{n}^{2}+(\frac{\Delta _{n}}{2})^{2}}$ and $|\varphi
_{\pm ,n}\rangle $, with $H_{\text{tot}}|\varphi _{\pm ,n}\rangle
=\varepsilon _{n,\pm }|\varphi _{\pm ,n}\rangle $. The gap between the
higher level of $n$-th pair and the lower level of $(n+1)$-th pair is $%
\varepsilon _{n+1,-}-\varepsilon _{n,+}=\omega _{n}-2\sqrt{J_{n}^{2}+(\frac{%
\Delta _{n}}{2})^{2}}$. We see that even for small $\Delta _{n}$ and $J_{n}$
(e.g., $\sqrt{J_{n}^{2}+(\frac{\Delta _{n}}{2})^{2}}\sim \frac{1}{4}\omega
_{n}$), the eigenenergy levels $\varepsilon _{n,\pm }$ would move by $\frac{1%
}{2}$ their spacing as shown in Fig.~3 in the main text.

The adiabaticity condition of the two-level Landau-Zener process is $\frac{%
\langle \varphi _{+,n}|\dot{\varphi}_{-,n}\rangle }{\varepsilon
_{n,+}-\varepsilon _{n,-}}\ll 1$. That is, the process duration should be
long compared to the inverse of the gap, leading to the speed proportional
to the gap. We emphasize that, the full description of the adiabatic process
is given by a multistate Landau-Zener problem by including off-resonance
couplings $(J_{n,m}|\varphi _{1,n}\rangle \langle \varphi _{2,m}|+h.c.)$.
The adiabaticity condition becomes $\frac{\langle \varphi _{\mp ,n}|\dot{%
\varphi}_{\pm ,m}\rangle }{\varepsilon _{n,\mp }-\varepsilon _{m,\pm }}\ll 1$%
, and the adiabatic duration is long compared to all the eigenenergy gaps $%
\varepsilon _{n,+}-\varepsilon _{n,-}$ and $\varepsilon _{n+1,-}-\varepsilon
_{n,+}$. Our full numerical simulation has taken into account all these
effects.

Since a larger gap leads to a faster qubit preparation, we would like to use
larger $J_{n}$ and $\Delta _{n}$ during the adiabatic process. As we discuss
above, $J_{n}$ and $\Delta _{n}$ should still be small compared to $\omega
_{n}$ to ensure that the all gaps are large enough. For example, one can use
$\sqrt{J_{n}^{2}+(\frac{\Delta _{n}}{2})^{2}}\sim \frac{1}{4}\omega _{n}$,
such that $(\varepsilon _{n,+}-\varepsilon _{n,-})\sim (\varepsilon
_{n+1,-}-\varepsilon _{n,+})\sim \frac{\omega _{n}}{2}$. Therefore, the
speed is limited by $\omega _{n}$. The speed can be very fast (10 ms or 100
ms preparation time for Li or Rb atoms) for the tweezers considered in this
paper.

\begin{figure}[t]
\includegraphics[width=1.0\linewidth]{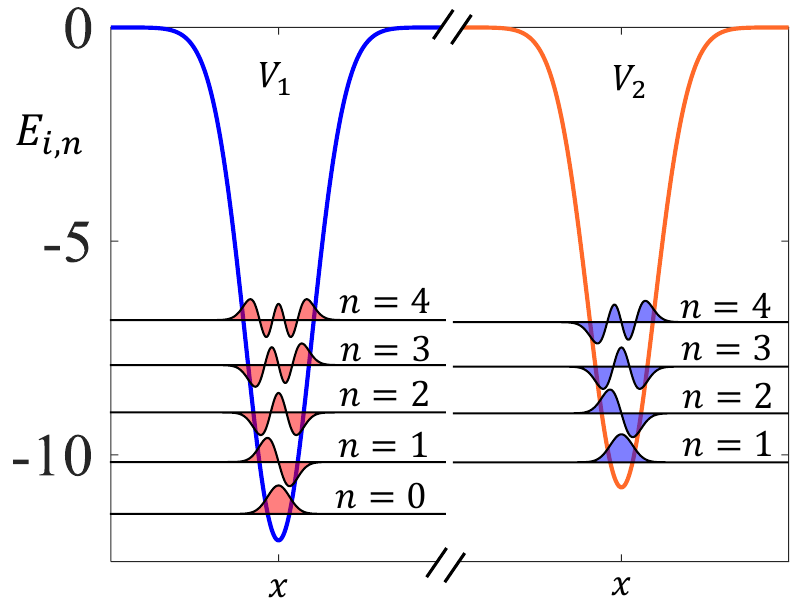}
\caption{The energy levels and corresponding wave functions for two Gaussian
traps satisfying approximate supersymmetry with $w_{0}=1\protect\mu $m, $%
\protect\alpha _{1}=-12E_{R}$ and $\protect\alpha _{2}=-10.76E_{R}$. The
results are obtained by numerically solving the Schr\"{o}dinger equation. $%
E_{R}$ is the energy unit.}
\label{fig:AS}
\end{figure}

\section{Appendix C: Approximate supersymmetry}

Exact supersymmetry requires isospectrality such that the corresponding
eigenvalues of the two tweezers are exactly matched initially [i.e., $%
E_{1,n}(t=0)=E_{2,n}(t=0)$]. In the presence of perturbations that weakly
break the energy degeneracy $E_{1,n}\neq E_{2,n}$ with $%
[E_{1,n}(t=0)-E_{2,n}(t=0)]\ll \omega _{n}(t=0)$, the supersymmetry becomes
an approximate symmetry (an approximate symmetry arises when the symmetry is
weakly broken). Our scheme works for both exact and approximate
supersymmetries, i.e., in the region $[E_{1,n}(t=0)-E_{2,n}(t=0)]\ll
\{J_{n}(t),\Delta _{n}(t)\}<\omega _{n}(t=0)$, such that all gaps remain
during the adiabatic process. If the supersymmetry is strongly broken, we
may have either unpaired states or extremely small gap during the
extraction. Therefore, the supersymmetry (either exact or approximate) is
crucial for the design of our tweezer geometry and plays a central role for
the atom extraction process. Two tweezer potentials that do not obey the
supersymmetry would not work.

Though it is possible to obtain exact supersymmetry by tailoring the tweezer
shapes, it is more realistic to work in the approximate supersymmetry
region. For the 1D Gaussian example, the auxiliary tweezer corresponds to an
approximate supersymmetric partner of the main tweezer (see Fig. \ref{fig:AS}%
). In particular, we are interested in the low-lying energy levels,
therefore, the traps $V_{1}(x)$ and $V_{2}(x)$ can be approximately
characterized by harmonic traps $V_{1}(x)\simeq \alpha
_{1}(1-2x^{2}/w_{0}^{2})$ and $V_{2}(x)\simeq \alpha
_{2}(1-2x^{2}/w_{0}^{2})\simeq V_{1}(x)+\Delta \alpha -2\Delta \alpha
x^{2}/w_{0}^{2}$. Besides the constant shift $\Delta \alpha \equiv (\alpha
_{2}-\alpha _{1})\ll \alpha _{1}$ compared to $V_{1}$, $V_{2}$ contains an
additional small term $2\Delta \alpha x^{2}/w_{0}^{2}$ which only leads to
slight differences in the energy splittings between the two traps, i.e., $%
|E_{1,n}-E_{2,n}|\simeq \frac{\Delta \alpha }{\alpha _{1}+\alpha _{2}}%
|E_{1,n+1}-E_{1,n}|$. For a constant shift equals to the energy splitting $%
\Delta \alpha =E_{1,n+1}-E_{1,n}$, we obtain $V_{2}$ as the approximate
superpartner of $V_{1}$ with $|E_{1,n}-E_{2,n}|\simeq \frac{1}{20}%
|E_{1,n+1}-E_{1,n}|$, as shown in Fig.~\ref{fig:AS}. Such tiny difference in
energy splitting can be suppressed further by slightly modifying the beam
waist of the auxiliary tweezer which eliminates the curvature difference $%
2\Delta \alpha x^{2}/w_{0}^{2}$. Similar results apply to the 3D tweezers.

\section{Appendix D: Difference between Boson and Fermion qubits}

We start from the single atom (several non-interacting atoms) in the tweezer
for bosonic (fermion) qubit. In both cases (single boson or several
fermions), the atom interaction is irrelevant in our ground state
preparation scheme because of the Pauli exclusion principle for fermions.

The tightly confined bosons strongly interact with each other, therefore
multi-occupation is avoided as one loads the reservoir atoms to the tweezer.
Through this method, defect-free single-atom tweezer arrays have been
experimentally realized by post-selecting and rearranging occupied tweezers~%
\cite{Science.354.1024,science.aah3778,Nature561.79,PhysRevLett.122.203601}.
The single atom after such process is hot, and further sideband cooling
could bring the ground state occupation probability to $\sim 90\%$, which
means still a few excited vibrational states could be occupied. Then one can
apply our supersymmetry cooling proposal to achieve ground state preparation
fidelity up to $\sim 99.99\%$.

The case for fermionic qubits is very different because two spinless
fermions do not occupy the same motional ground state and interact with each
other. Therefore the method for bosonic qubit does not apply for fermion
qubit. A different laser culling method~\cite{PhysRevA.80.030302} can be
used to prepare a few atoms in an optical tweezer from a reservoir of
degenerate fermi gas through gradually reducing the optical tweezer
potential, as demonstrated in experiments~\cite{science.1201351}. In this
case, the ground state is already occupied with a high probability, and the
difficulty is how to remove the last few low-lying excited state atoms
without affecting the ground state atom. In our fermion qubit preparation,
we start from a few spinless atoms distributed on the low-lying energy
levels, and the supersymmetry scheme can extract all excited fermions out.
These spinless atoms have no interaction.



\begin{thebibliography}{99}
\bibitem{Schlosser2001sub} N.~Schlosser, G.~Reymond, I.~Protsenko, and
P.~Grangier, Sub-poissonian loading of single atoms in a microscopic dipole
trap, \href{https://doi.org/10.1038/35082512}{Nature (London) \textbf{411},
1024 (2001)}.

\bibitem{RevModPhys.82.2313} M.~Saffman, T.~G.~Walker, and K.~M\"{o}lmer,
Quantum information with Rydberg atoms, \href{https://doi.org/10.1103/RevModPhys.82.2313}%
{Rev. Mod. Phys. \textbf{82}, 2313 (2010)}.

\bibitem{RevModPhys.86.153} I.~M.~Georgescu, S.~Ashhab, and F.~Nori, Quantum
simulation, \href{https://doi.org/10.1103/RevModPhys.86.153}{Rev. Mod. Phys.
\textbf{86}, 153 (2014)}.

\bibitem{PhysToday.70.44} D.~S.~Weiss and M.~Saffman, Quantum computing with
neutral atoms, \href{https://doi.org/10.1063/PT.3.3626}{Phys. Today \textbf{%
70}, 44 (2017)}.

\bibitem{Science.357.995} C.~Gross and I.~Bloch, Quantum simulations with
ultracold atoms in optical lattices, \href{https://doi.org/10.1126/science.aal3837}%
{Science \textbf{357}, 995 (2017)}.

\bibitem{PhysRevA.70.040302} D. S. Weiss, J. Vala, A. V. Thapliyal, S.
Myrgren, U. Vazirani, and K. B. Whaley, Another way to approach zero entropy
for a finite system of atoms, \href{https://doi.org/10.1103/PhysRevA.70.040302}%
{Phys. Rev. A \textbf{70}, 040302R (2004)}.

\bibitem{Science.354.1024} M. Endres, H. Bernien, A. Keesling, H. Levine, E.
R. Anschuetz, A. Krajenbrink, C. Senko, V. Vuletic, M. Greiner, and M. D.
Lukin, Atom-by-atom assembly of defect-free one-dimensional cold atom
arrays, \href{https://doi.org/10.1126/science.aah3752}{Science \textbf{354},
1024 (2016)}.

\bibitem{ncomms13317} H.~Kim, W.~Lee, H.~Lee, H.~Jo, Y.~Song, and J.~Ahn,
\emph{In situ} single-atom array synthesis using dynamic holographic optical
tweezers, \href{https://doi.org/10.1038/ncomms13317 (2016)}{Nat. Commun.
\textbf{7}, 13317 (2016)}.

\bibitem{science.aah3778} D.~Barredo, S.~de L{\'e}s{\'e}leuc, V.~Lienhard,
T.~Lahaye, and A.~Browaeys, An atom-by-atom assembler of defect-free
arbitrary two-dimensional atomic arrays, \href{https://doi.org/10.1126/science.aah3778}%
{Science, \textbf{354} 1021 (2016)}.

\bibitem{OE.24.009816} W. Lee, H. Kim, and J. Ahn, Three-dimensional
rearrangement of single atoms using actively controlled optical microtraps,
\href{https://doi.org/10.1364/OE.24.009816}{Opt. Express \textbf{24}, 9816
(2016)}.

\bibitem{Nature561.79} D. Barredo, V. Lienhard, S. de L{\'e}s{\'e}leuc, T.
Lahaye, and A.~Browaeys, Synthetic three-dimensional atomic structures
assembled atom by atom, \href{https://doi.org/10.1038/s41586-018-0450-2}{%
Nature (London) \textbf{561}, 79 (2018)}.

\bibitem{PhysRevLett.122.203601} D.~O.~de Mello, D.~Sch\"affner,
J.~Werkmann, T.~Preuschoff, L.~Kohfahl, M.~Schlosser, and G.~Birkl,
Defect-Free Assembly of 2D Clusters of More Than 100 Single-Atom Quantum
Systems, \href{https://doi.org/10.1103/PhysRevLett.122.203601}{Phys. Rev.
Lett. \textbf{122}, 203601 (2019)}.

\bibitem{PhysRevLett.96.063001} D. D. Yavuz, P. B. Kulatunga, E. Urban, T.
A. Johnson, N. Proite, T. Henage, T. G. Walker, and M. Saffman, Fast Ground
State Manipulation of Neutral Atoms in Microscopic Optical Traps, \href{https://doi.org/10.1103/PhysRevLett.96.063001}%
{Phys. Rev. Lett. \textbf{96}, 063001 (2006)}.

\bibitem{PhysRevA.74.042316} C.~Zhang, S. L. Rolston, and S. Das Sarma,
Manipulation of single neutral atoms in optical lattices, \href{https://doi.org/10.1103/PhysRevA.74.042316}%
{Phys. Rev. A \textbf{74}, 042316 (2006)}.

\bibitem{nature09827} C.~Weitenberg, M.~Endres, J.~F.~Sherson, M.~Cheneau,
P.~Schau\ss , T.~Fukuhara, I.~Bloch, and S.~Kuhr, Single-spin addressing in
an atomic Mott insulator, \href{https://doi.org/10.1038/nature09827}{Nature
\textbf{471}, 319 (2011)}.

\bibitem{PhysRevLett.93.150501} D. Schrader, I. Dotsenko, M. Khudaverdyan,
Y. Miroshnychenko, A. Rauschenbeutel, and D. Meschede, Neutral Atom Quantum
Register, \href{https://doi.org/10.1103/PhysRevLett.93.150501}{Phys. Rev.
Lett. \textbf{93}, 150501 (2004)}.

\bibitem{PhysRevA.77.052309} T. R. Beals, J. Vala, and K. B. Whaley,
Scalability of quantum computation with addressable optical lattices, \href{https://doi.org/10.1103/PhysRevA.77.052309}%
{Phys. Rev. A \textbf{77}, 052309 (2008)}.

\bibitem{science.aaf2581} Y.~Wang, A.~Kumar, T.-Y.~Wu, and D.~S.~Weiss,
Single-qubit gates based on targeted phase shifts in a 3D neutral atom
array, \href{https://doi.org/10.1126/science.aaf2581}{Science \textbf{352},
1562 (2016)}.

\bibitem{PhysRevLett.121.240501} C. Sheng, X. He, P. Xu, R. Guo, K. Wang, Z.
Xiong, M. Liu, J. Wang, and M. Zhan, High-Fidelity Single-Qubit Gates on
Neutral Atoms in a Two-Dimensional Magic-Intensity Optical Dipole Trap
Array, \href{https://doi.org/10.1103/PhysRevLett.121.240501}{Phys. Rev.
Lett. \textbf{121}, 240501 (2018)}.

\bibitem{Nature527.208} A. M. Kaufman, B. J. Lester, M. Foss-Feig, M. L.
Wall, A. M. Rey, and C. A. Regal, Entangling two transportable neutral atoms
via local spin exchange, \href{https://doi.org/10.1038/nature16073}{Nature
(London) \textbf{527}, 208 (2015)}.

\bibitem{PhysRevLett.85.2208} D. Jaksch, J. I. Cirac, P. Zoller, S. L.
Rolston, R. C{\^o}t{\'e}, and M. D. Lukin, Fast Quantum Gates for Neutral
Atoms, \href{https://doi.org/10.1103/PhysRevLett.85.2208}{Phys. Rev. Lett.
\textbf{85}, 2208 (2000)}.

\bibitem{PhysRevLett.104.010503} L. Isenhower, E. Urban, X. L. Zhang, A. T.
Gill, T. Henage, T. A. Johnson, T. G. Walker, and M. Saffman, Demonstration
of a Neutral Atom Controlled-NOT Quantum Gate, \href{https://doi.org/10.1103/PhysRevLett.104.010503}%
{Phys. Rev. Lett. \textbf{104}, 010503 (2010)}.

\bibitem{PhysRevLett.104.010502} T. Wilk, A. Ga\"{e}tan, C. Evellin, J.
Wolters, Y. Miroshnychenko, P. Grangier, and A. Browaeys, Entanglement of
Two Individual Neutral Atoms Using Rydberg Blockade, \href{https://doi.org/10.1103/PhysRevLett.104.010502}%
{Phys. Rev. Lett. \textbf{104}, 010502 (2010)}.

\bibitem{nphys3487} Y. Y. Jau, A. M. Hankin, T. Keating, I. H. Deutsch, and
G. W. Biedermann, Entangling atomic spins with a Rydberg-dressed spin-flip
blockade, \href{https://doi.org/10.1038/nphys3487}{Nat. Phys. \textbf{12},
71 (2016)}.

\bibitem{nature24622} H. Bernien, S. Schwartz, A. Keesling, H. Levine, A.
Omran, H. Pichler, S. Choi, A. S. Zibrov, M. Endres, M. Greiner, V. Vuletic,
and M. D. Lukin, Probing many-body dynamics on a 51-atom quantum simulator,
\href{https://doi.org/10.1038/nature24622}{Nature \textbf{551}, 579 (2017)}.

\bibitem{PhysRevLett.121.123603} H.~Levine, A.~Keesling, A.~Omran,
H.~Bernien, S.~Schwartz, A.~S.~Zibrov, M.~Endres, M.~Greiner, V.~Vuleti%
\ifmmode \acute{c}\else \'{c}\fi{}, and M.~D.~Lukin, High-Fidelity Control
and Entanglement of Rydberg-Atom Qubits, \href{https://doi.org/10.1103/PhysRevLett.121.123603}%
{Phys. Rev. Lett. \textbf{121}, 123603 (2018)}.

\bibitem{PhysRevX.8.021070} V. Lienhard, S. de L\'{e}s\'{e}leuc, D. Barredo,
T. Lahaye, A. Browaeys, M. Schuler, L.-P. Henry, and A. M. L\"{a}uchli,
Observing the Space- and Time-Dependent Growth of Correlations in
Dynamically Tuned Synthetic Ising Models with Antiferromagnetic
Interactions, \href{https://doi.org/10.1103/PhysRevX.8.021070}{Phys. Rev. X
\textbf{8}, 021070 (2019)}.

\bibitem{Saffman2019} T.\thinspace M. Graham, M. Kwon, B. Grinkemeyer, Z.
Marra, X. Jiang, M.\thinspace T. Lichtman, Y. Sun, M. Ebert, and M. Saffman,
Rydberg-Mediated Entanglement in a Two-Dimensional Neutral Atom Qubit Array,
\href{https://doi.org/10.1103/PhysRevLett.123.230501}{Phys. Rev. Lett.
\textbf{123}, 230501 (2019)}.

\bibitem{PhysRevX.2.041014} A. M. Kaufman, B. J. Lester, C. A. Regal,
Cooling a single atom in an optical tweezer to its quantum ground state,
\href{https://doi.org/10.1103/PhysRevX.2.041014}{Phys. Rev. X \textbf{2},
041014 (2012)}.

\bibitem{PhysRevLett.110.133001} J. D. Thompson, T. G. Tiecke, A. S. Zibrov,
V. Vuleti\ifmmode \acute{c}\else \'{c}\fi{}, and M. D. Lukin, Coherence and
Raman Sideband Cooling of a Single Atom in an Optical Tweezer, \href{https://doi.org/10.1103/PhysRevLett.110.133001}%
{Phys. Rev. Lett. \textbf{110}, 133001 (2013)}.

\bibitem{PhysRevA.97.063423} Y. Yu, N. R. Hutzler, J. T. Zhang, L. R. Liu,
J. D. Hood, T. Rosenband, and K.-K. Ni, Motional-ground-state cooling
outside the Lamb-Dicke regime, \href{https://doi.org/10.1103/PhysRevA.97.063423}%
{Phys. Rev. A \textbf{97}, 063423 (2018)}.

\bibitem{PhysRevX.9.021039} L.~R.~Liu, J.~D.~Hood, Y.~Yu, J.~T.~Zhang,
K.~Wang, Y.-W.~Lin, T.~Rosenband, and K.-K.~Ni, Molecular Assembly of
Ground-State Cooled Single Atoms, \href{https://doi.org/10.1103/PhysRevX.9.021039}%
{Phys. Rev. X \textbf{9}, 021039 (2019)}.

\bibitem{PhysRevX.8.041055} A.~Cooper, J.~P.~Covey, I.~S.~Madjarov,
S.~G.~Porsev, M.~S.~Safronova, and M.~Endres, Alkaline-Earth Atoms in
Optical Tweezers, \href{https://doi.org/10.1103/PhysRevX.8.041055}{Phys.
Rev. X \textbf{8}, 041055 (2018)}.

\bibitem{PhysRevX.8.041054} M.~A.~Norcia, A.~W.~Young, and A.~M.~Kaufman,
Microscopic Control and Detection of Ultracold Strontium in Optical-Tweezer
Arrays, \href{https://doi.org/10.1103/PhysRevX.8.041054}{Phys. Rev. X
\textbf{8}, 041054 (2018)}.

\bibitem{PhysRevLett.122.143002} S.~Saskin, J.~T.~Wilson, B.~Grinkemeyer,
and J.~D.~Thompson, Narrow-Line Cooling and Imaging of Ytterbium Atoms in an
Optical Tweezer Array, \href{https://doi.org/10.1103/PhysRevLett.122.143002}{%
Phys. Rev. Lett. \textbf{122}, 143002 (2019)}.

\bibitem{PhysRevA.80.030302} M.~G.~Raizen, S.-P.~Wan, C.~Zhang, and Q.~Niu,
Ultrahigh-fidelity qubits for quantum computing, \href{https://doi.org/10.1103/PhysRevA.80.030302}%
{Phys. Rev. A \textbf{80}, 030302 (2009)}.

\bibitem{science.1201351} F. Serwane, G. Z\"urn, T. Lompe, T. B. Ottenstein,
A. N. Wenz, and S. Jochim, Deterministic Preparation of a Tunable
Few-Fermion System, \href{https://doi.org/10.1126/science.1201351}{Science,
\textbf{332}, 336 (2011)}.

\bibitem{science.1240516} A.~N.~Wenz, G.~Z\"urn, S.~Murmann, I.~Brouzos,
T.~Lompe, and S.~Jochim, From Few to Many: Observing the Formation of a
Fermi Sea One Atom at a Time, \href{https://doi.org/10.1126/science.1240516}{%
Science, \textbf{342}, 457 (2013)}.

\bibitem{PhysRevLett.115.215301} S.~Murmann, F.~Deuretzbacher, G.~Z\"urn,
J.~Bjerlin, S.~M.~Reimann, L.~Santos, T.~Lompe, and S.~Jochim,
Antiferromagnetic Heisenberg Spin Chain of a Few Cold Atoms in a
One-Dimensional Trap, \href{https://doi.org/10.1103/PhysRevLett.115.215301}{%
Phys. Rev. Lett. \textbf{115}, 215301 (2015)}.

\bibitem{Supersymmetry2015} M.~Dine, Supersymmetry and string theory: Beyond
the standard model (Cambridge University Press, 2015).

\bibitem{Supersymmetry1995Cooper} F.~Cooper, A.~Khare, and U.~Sukhatme,
Supersymmetry and quantum mechanics, \href{https://doi.org/10.1016/0370-1573(94)00080-M}%
{Phys. Rep. \textbf{251}, 267 (1995)}.

\bibitem{PhysRevLett.100.090404} Y.~Yu and K.~Yang, Supersymmetry and the
Goldstino-Like Mode in Bose-Fermi Mixtures, \href{https://doi.org/10.1103/PhysRevLett.100.090404}%
{Phys. Rev. Lett. \textbf{100}, 090404 (2008)}.

\bibitem{PhysRevLett.110.233902} M.-A.~Miri, M.~Heinrich, R.~El-Ganainy, and
D.~N.~Christodoulides, Supersymmetric Optical Structures, \href{https://doi.org/10.1103/PhysRevLett.110.233902}%
{Phys. Rev. Lett. \textbf{110}, 233902 (2013)}.

\bibitem{ncomms4698} M. Heinrich, M.-A. Miri, S. St\"utzer, R. El-Ganainy,
S. Nolte, A. Szameit, and D. N. Christodoulides, Supersymmetric mode
converters, \href{https://doi.org/10.1038/ncomms4698}{Nat. Commun. \textbf{5}%
, 3698 (2014)}.

\bibitem{OL.43.004927} B. Midya, W. Walasik, N. M. Litchinitser, and L.
Feng, Supercharge optical arrays, \href{https://doi.org/10.1364/OL.43.004927}%
{Opt. Lett. \textbf{43}, 4927 (2018)}.

\bibitem{LiangFeng2019Super} B.~Midya, H.~Zhao, X.~Qiao, P.~Miao,
W.~Walasik, Z.~Zhang, N.~M.~Litchinitser, and L.~Feng, Supersymmetric
microring laser arrays, \href{https://doi.org/10.1364/PRJ.7.000363}{%
Photonics Research \textbf{7}, 363 (2019)}.

\bibitem{science.aav5103} M.~P.~Hokmabadi, N.~S.~Nye, R.~El-Ganainy,
D.~N.~Christodoulides, and M.~Khajavikhan, Supersymmetric laser arrays,
\href{https://doi.org/10.1126/science.aav5103}{Science, \textbf{363}, 623
(2019)}.


\bibitem{PhysRevLett.86.1514} V.~Milner, J.~L.~Hanssen, W.~C.~Campbell, and
M.~G.~Raizen, Optical Billiards for Atoms, \href{https://doi.org/10.1103/PhysRevLett.86.1514}%
{Phys. Rev. Lett. \textbf{86}, 1514 (2001)}.

\bibitem{PhysRevLett.86.1518} N.~Friedman, A.~Kaplan, D.~Carasso, and
N.~Davidson, Observation of Chaotic and Regular Dynamics in Atom-Optics
Billiards, \href{https://doi.org/10.1103/PhysRevLett.86.1518}{Phys. Rev.
Lett. \textbf{86}, 1518 (2001)}.


\bibitem{J.Mod.Opt.54.1619} M.~Schulz, H.~Crepaz, F.~Schmidt-Kaler,
J.~Eschner, and R.~Blatt, Transfer of trapped atoms between two optical
tweezer potentials, \href{http://dx.doi.org/10.1080/09500340600861740}{J.
Mod. Opt. 54, 1619 (2007)}.

\bibitem{nature09378} J.~F.~Sherson, C.~Weitenberg, M.~Endres, M.~Cheneau,
I.~Bloch, and S.~Kuhr, Single-atom-resolved fluorescence imaging of an
atomic Mott insulator, \href{https://doi.org/10.1038/nature09378}{Nature
\textbf{467}, 68 (2010)}.


\bibitem{PhysRevA.63.033603} L. Viverit, S. Giorgini, L. P. Pitaevskii, and
S. Stringari, Adiabatic compression of a trapped Fermi gas, \href{https://doi.org/10.1103/PhysRevA.63.033603}%
{Phys. Rev. A \textbf{63}, 033603 (2001)}.

\bibitem{AdvAtMolOptPhys.42.95} R.~Grimm, M.~Weidem\"uller, and
Y.~B.~Ovchinnikov, Optical Dipole Traps for Neutral Atoms, \href{https://doi.org/10.1016/S1049-250X(08)60186-X}%
{Adv. At. Mol. Opt. Phys. \textbf{42}, 95 (2000)}.

\bibitem{OptLett.34.2912} P.~Kwee, B.~Willke, and K.~Danzmann,
Shot-noise-limited laser power stabilization with a high-power photodiode
array, \href{https://doi.org/10.1364/OL.34.002912}{Opt. Lett. \textbf{34},
2912 (2009)}.

\bibitem{OptLett.42.755} J.~Junker, P.~Oppermann, and B.~Willke,
Shot-noise-limited laser power stabilization for the AEI 10 m Prototype
interferometer, \href{https://doi.org/10.1364/OL.42.000755}{Opt. Lett.
\textbf{42}, 755 (2017)}.

\bibitem{PhysRevLett.121.173601} H.~Vahlbruch, D.~Wilken, M.~Mehmet, and
B.~Willke, Laser Power Stabilization beyond the Shot Noise Limit Using
Squeezed Light, \href{https://doi.org/10.1103/PhysRevLett.121.173601}{Phys.
Rev. Lett. \textbf{121}, 173601 (2018)}.

\bibitem{PhysRevA.56.R1095} T.~A.~Savard, K.~M.~O'Hara, and J.~E.~Thomas,
Laser-noise-induced heating in far-off resonance optical traps, \href{https://doi.org/10.1103/PhysRevA.56.R1095}%
{Phys. Rev. A \textbf{56}, R1095(R) (1997)}.

\bibitem{PhysRevA.80.032307} C.-S.~Chuu and C.~Zhang, Suppression of phase
decoherence in a single atomic qubit, \href{https://doi.org/10.1103/PhysRevA.80.032307}%
{Phys. Rev. A \textbf{80}, 032307 (2009)}.

\bibitem{PhysRevLett.75.4011} C.~Monroe, D.~M.~Meekhof, B.~E.~King,
S.~R.~Jefferts, W.~M.~Itano, D.~J.~Wineland, and P.~Gould, Resolved-Sideband
Raman Cooling of a Bound Atom to the 3D Zero-Point Energy, \href{https://doi.org/10.1103/PhysRevLett.75.4011}%
{Phys. Rev. Lett. \textbf{75}, 4011 (1995)}.

\bibitem{RevModPhys.90.035005} L.~Pezz\`e, A.~Smerzi, M.~K.~Oberthaler,
R.~Schmied, and P.~Treutlein, Quantum metrology with nonclassical states of
atomic ensembles, \href{https://doi.org/10.1103/RevModPhys.90.035005}{Rev.
Mod. Phys. \textbf{90}, 035005 (2018)}.

\bibitem{PhysRevLett.91.010402} A.~M.~Dudarev, R.~B.~Diener, B.~Wu,
M.~G.~Raizen, and Q.~Niu, Entanglement Generation and Multiparticle
Interferometry with Neutral Atoms, \href{https://doi.org/10.1103/PhysRevLett.91.010402}%
{Phys. Rev. Lett. \textbf{91}, 010402 (2003)}.

\bibitem{Science.326.1683} S.~E.~Pollack, D.~Dries, and R.~G.~Hulet,
Universality in Three- and Four-Body Bound States of Ultracold Atoms, \href{https://doi.org/10.1126/science.1182840}%
{Science \textbf{326}, 1683 (2009)}.

\bibitem{Rep.Prog.Phys.75.046401} D.~Blume, Few-body physics with ultracold
atomic and molecular systems in traps, \href{https://doi.org/10.1088/0034-4885/75/4/046401}%
{Rep. Prog. Phys. \textbf{75}, 046401 (2012)}.

\bibitem{Rep.Prog.Phys.80.056001} P.~Naidon, S.~Endo, Efimov Physics: a
review, \href{https://doi.org/10.1088/1361-6633/aa50e8}{Rep. Prog. Phys.
\textbf{80}, 056001 (2017)}.

\bibitem{RevModPhys.89.035006} C.~H.~Greene, P.~Giannakeas, and J.~P\'erez-R%
\'{\i}os, Universal few-body physics and cluster formation, \href{https://doi.org/10.1103/RevModPhys.89.035006}%
{Rev. Mod. Phys. \textbf{89}, 035006 (2017)}.
\end{thebibliography}


\end{document}